\documentclass[a4paper,12pt]{article}

\newcommand{\sect}[1]{\setcounter{equation}{0}\section{#1}}

\begin{document}
\topmargin 0pt
\oddsidemargin 0mm

\renewcommand{\thefootnote}{\fnsymbol{footnote}}
\begin{titlepage}
\begin{flushright}
OU-HET 341 \\
hep-th/0001213
\end{flushright}

\vspace{5mm}
\begin{center}
{\Large \bf Noncommutative and Ordinary Super Yang-Mills on (D$(p-2)$, D$p$)
Bound States}
\vspace{10mm}

{\large
Rong-Gen Cai\footnote{e-mail address: cai@het.phys.sci.osaka-u.ac.jp} and
Nobuyoshi Ohta\footnote{e-mail address: ohta@phys.sci.osaka-u.ac.jp}} \\
\vspace{8mm}
{\em Department of Physics, Osaka University,
Toyonaka, Osaka 560-0043, Japan}

\end{center}
\vspace{5mm}
\centerline{{\bf{Abstract}}}
\vspace{5mm}

We study properties of (D$(p-2)$, D$p$) nonthreshold bound states
($2 \le p \le 6 $) in the dual gravity description. These bound
states can be viewed as D$p$-branes with a nonzero NS $B$ field of rank
two. We find that in the decoupling limit, the thermodynamics of the $N_p$
coincident D$p$-branes with $B$ field is the same not only as that of $N_p$
coincident D$p$-branes without $B$ field, but also as that of the $N_{p-2}$
coincident D$(p-2)$-branes with two smeared coordinates and no $B$ field,
for $N_{p-2}/N_p= \tilde{V}_2/[(2\pi)^2 \tilde{b}]$ with $\tilde{V}_2$ being
the area of the two smeared directions and $\tilde{b}$ a noncommutativity
parameter. We also obtain the same relation from the thermodynamics and
dynamics by probe methods. This suggests that the noncommutative super
Yang-Mills with gauge group $U(N_p)$ in ($p+1$) dimensions is equivalent to
an ordinary one with gauge group $U(\infty)$ in ($p-1$) dimensions in the
limit $\tilde{V}_2 \to \infty$. We also find that the free energy
of a D$p$-brane probe with $B$ field in the background of D$p$-branes with
$B$ field coincides with that of a D$p$-brane probe in the background of
D$p$-branes without $B$ field.

\end{titlepage}

\newpage
\renewcommand{\thefootnote}{\arabic{footnote}}
\setcounter{footnote}{0}
\setcounter{page}{2}

\sect{Introduction}

Recently there has been much interest in studying the properties of
the (D$(p-2)$, D$p$) bound states and their consequences (see for
example~\cite{Hashimoto1}-\cite{Lu} and references therein).
Such bound states can be viewed as D$p$-branes with a nonzero (rank two)
Neveu-Schwarz (NS) $B$ field. At present, it is known that the worldvolume
coordinates will become noncommutative if a D$p$-brane carries a
nonvanishing NS $B$ field on its worldvolume~\cite{Li}-\cite{Seiberg},
and the field theories on the worldvolume of such D-branes are called
noncommutative field theories in order to distinguish the ordinary field
theories on the worldvolume of D-branes without NS $B$ field. The gauge
theories on the noncommutative spacetimes can naturally be realized in
string theories. According to the Maldacena
conjecture~\cite{Mald2}-\cite{Itzhaki}, the string theories can be used to
study the large $N$ noncommutative field theories in the strong 't Hooft
coupling limit.

On the basis of consideration of planar diagrams, Bigatti and Susskind
\cite{BS} argued that the large $N$ noncommutative and ordinary gauge field
theories are equivalent in the weak coupling limit and noncommutative
effects can be seen only in the nonplanar diagrams. Explicit perturbative
calculations~\cite{Arcioni} render evidence to this assertion. On the
supergravity side, it has also been found that the thermodynamics of
near-extremal D$p$-branes with a nonvanishing NS $B$ field coincides exactly
with that of the corresponding D$p$-branes without $B$
field~\cite{Mald1,Ali,Bar,CO,Harmark}, which means that in the large $N$
and strong coupling limit, the number of the degrees of freedom of
the noncommutative gauge fields remains unchanged despite the
noncommutativity of space.

More recently, Lu and Roy~\cite{Lu} have found that in the system of
(D$(p-2)$, D$p$) bound states ($2 \le p \le 6$), the noncommutative effects
of gauge fields are actually due to the presence of {\it infinitely many}
D$(p-2)$-branes which play the dominant role over the D$p$-branes in the
large $B$ field limit. The D$p$-branes with a constant $B$ field represent
dynamically the system of infinitely many D$(p-2)$-branes with two smeared
transverse coordinates (additional isometrics) and no $B$ field in the
decoupling limit. With this observation, Lu and Roy further argued that
there is an equivalence between the noncommutative super Yang-Mills theory
in $(p+1)$ dimensions and an ordinary one with gauge group $U(\infty)$ in
$(p-1)$ dimensions. For related discussions, see
also~\cite{T,Ishibashi1,Ishibashi2,CS,Corn,Kato}.

In the present paper, we would like to discuss this equivalence from the
viewpoint of the thermodynamics of the (D$(p-2)$, D$p$) bound states
in the dual gravity description with two dimensions compactified on a torus.
The case discussed by Lu and Roy~\cite{Lu} corresponds to the infinite
volume limit of the torus, and we would like to clarify some subtle questions
in this analysis.

In the next section, we study the black (D$(p-2)$, D$p$) configuration and
some of its basic thermodynamic properties. As mentioned above, it has been
noticed that the thermodynamics of the black D$p$-branes with $B$ field is
the same as that of D$p$-branes without $B$ field. We show here that the
thermodynamics of the D$p$-branes with nonzero $B$ field is also completely
the same as that of D$(p-2)$-branes with two smeared coordinates and zero
$B$ field. We obtain a relation [eq.~(\ref{number}) below] between
the numbers of D$p$-branes with $B$ field and the D$(p-2)$-branes without
$B$ field when this equivalence is valid. Since the worldvolume theory of
D$p$-branes with $B$ field is the noncommutative super Yang-Mills with gauge
group $U(N_p)$ in ($p+1$) dimensions with two dimensions compactified on
a torus and that of D$(p-2)$-branes with two smeared coordinates and zero $B$
field is an ordinary super Yang-Mills with gauge group $U(N_{p-2})$ in ($p+1$)
dimensions on the dual torus, as we will see, this implies that these
theories are equivalent in the large $N$ limit. When the volume of the torus
is sent to infinity, the latter theory reduces to $U(\infty)$ gauge theory
in ($p-1$) dimensions for fixed $N_p$, in agreement with \cite{Lu}.
This is also in accordance with the proposal that D2-brane is a condensate
of D0-branes~\cite{T,Ishibashi1,Ishibashi2}. Furthermore, we give the
relation of Yang-Mills coupling constants between the theories in different
dimensions.

Investigating the interactions between a probe and a source is a useful
method to see some of the properties of the source. For instance, a scalar
field has been used~\cite{Kaya,Myung1,Myung2} as a probe to find out the
noncommutative effects on the absorption by the (D1, D3) bound state.
In section 3, we study the descriptions of the (D$(p-2)$, D$p$) bound
state in terms of the D$p$-branes with $B$ field and in terms of the
D$(p-2)$-branes without $B$ field by analyzing the thermodynamics of two
probes, one of which is a bound state of D$(p-2)$- and D$p$-branes and the
other is a D$(p-2)$-brane. In the decoupling limit, we find a relation
(\ref{3e18}) similar to (\ref{number}). In section 4, we further
reveal the equivalence of the descriptions by examining the dynamics of
the probes in the bound state backgrounds. We find that non-extremal
D-branes can be located at the horizon from the viewpoint of probes.
We summarize our results in section 5.

\sect{The (D$(p-2)$, D$p$) bound states and implications
 of their thermodynamics }

The supergravity solutions of the (D$(p-2)$, D$p$) bound states
($2 \le p \le 6$ ) in type II superstring theories have been constructed
by many authors in \cite{Russo,Breck,Mig,Roy,Harmark}. The supergravity
solutions of the (D$(p-2)$, D$p$) bound states can be acquired by taking
the extremal limit of the corresponding black configurations.

We start with the general solution
\begin{eqnarray}
&& ds^2 = H^{-1/2}[-f dt^2 +dx_1^2 +\cdots +dx_{p-2}^2 +h(dx_{p-1}^2 +dx_p^2)]
 +H^{1/2}(f^{-1}dr^2 +r^2 d\Omega_{8-p}^2), \nonumber\\
&& e^{2\phi} = g^2 H^{\frac{3-p}{2}}h, \ \ \ B_{p-1,p}=\tan\theta H^{-1}h,
 \nonumber\\
\label{2e1}
&& A^{p}_{012\cdots p} = g^{-1}(H^{-1}-1)h\cos\theta \coth\alpha,\ \ \
 A^{p-2}_{012\cdots (p-2)}=g^{-1}(H^{-1}-1)\sin\theta \coth\alpha,
\end{eqnarray}
and
\begin{equation}
\label{2e2}
H=1+\frac{r_0^{7-p}\sinh^2\alpha}{r^{7-p}}, \ \ \ f=1-\left(\frac{r_0}
 {r}\right)^{7-p}, \ \ \ h^{-1}=\cos^2\theta +H^{-1}\sin^2\theta.
\end{equation}
Here $g$ is the string coupling constant, $r_0$ is the non-extremal
Schwarzschild mass parameter and $\alpha $ is the boost parameter. The
solution (\ref{2e1}) interpolates between the black D$(p-2)$-brane solution
with two smeared coordinates $x_{p-1}$ and $x_p$ ($\theta=\pi/2$), and
the black D$p$-brane with zero $B$ field ($\theta=0$).
Note that a constant part of the $B$ field can be gauged away so that the
constant value for $B_{p-1,p}$ can be changed. The parameter
$\theta$ characterizes the interpolation. The coordinates $x_{p-1}$ and $x_p$
are relative transverse directions for the D$(p-2)$-branes and parametrize
a rectangular 2-torus.

Denote the area of the 2-torus spanned by $x_{p-1}$ and $x_p$ by $V_2$ and
the spatial volume of the D$(p-2)$-brane with worldvolume coordinates $(t,
x_1, \cdots, x_{p-2})$ by $V_{p-2}$. The spatial volume of the D$p$-brane
with worldvolume coordinates $(t, x_1, \cdots, x_p)$ is then
$V_p=V_{p-2}V_2$. The charge density of the D$p$-brane in the bound
state system is given by
\begin{equation}
\label{2e3}
Q_p =\frac{1}{2\kappa^2}\int_{\Omega_{8-p}}*F_{p+2}
 =\frac{(7-p)\Omega_{8-p} \cos\theta}{2\kappa^2 g}r_0^{7-p}
  \sinh\alpha \cosh\alpha,
\end{equation}
where $2\kappa^2 =(2\pi)^7 \alpha'^4$ is the gravity constant in ten
dimensions and $\Omega_{8-p}$ is the volume of a unit $(8-p)$-sphere:
\begin{equation}
\Omega_{8-p}=\frac{2\pi^{(9-p)/2}}{\Gamma[(9-p)/2]}=\frac{4\pi\cdot
  \pi ^{(7-p)/2}}{(7-p)\Gamma[(7-p)/2]}.
\end{equation}
The D$(p-2)$-brane charge density on its worldvolume is
\begin{equation}
Q_{p-2}=\frac{1}{2\kappa^2}\int_{V_2\times\Omega_{8-p}}*F_p
    =\frac{(7-p)\Omega_{8-p}V_2 \sin \theta}{2\kappa^2 g}
      r_0^{7-p}\sinh\alpha\cosh\alpha .
\label{2e5}
\end{equation}
In fact the D$(p-2)$-brane charge density on the worldvolume of D$p$-brane is
\begin{equation}
\tilde{Q}_{p-2}=\frac{Q_{p-2}}{V_2}.
\end{equation}
According to the charge quantization rule $Q_p=T_pN_p$ in terms of the
tension of the D$p$-branes, we can obtain the number $N_p$ of the D$p$-branes
and $N_{p-2}$ of the D$(p-2)$-branes in the bound states. Defining
$\tilde{R}^{7-p}=r_0^{7-p}\sinh\alpha\cosh\alpha$, we have the relation
between the number $N_p$ of D$p$-branes and $N_{p-2}$ of D$(p-2)$-branes:
\begin{equation}
\tilde{R}^{7-p} = N_p\frac{2\kappa^2 gT_p}{(7-p)\Omega_{8-p}\cos\theta}
 =N_{p-2}\frac{2\kappa^2 gT_{p-2}}{(7-p)\Omega_{8-p}V_2\sin\theta},
\end{equation}
where the tensions $T_p$ and $T_{p-2}$ have a unified expression as
$T_p=(2\pi)^{-p}(\alpha')^{-(p+1)/2}$.

{}From (\ref{2e3}) and (\ref{2e5}), we can see that the asymptotic
value $\tan \theta $ of the $B$ field has the following relation to the
charges of the D$p$- and D$(p-2)$-branes:
\begin{equation}
\label{2e7}
\tan\theta =\frac{\tilde{Q}_{p-2}}{Q_p}=\frac{1}{V_2}\frac{Q_{p-2}}{Q_p}
=\frac{1}{V_2}\frac{T_{p-2}}{T_p}\frac{N_{p-2}}{N_p}.
\end{equation}

The solution (\ref{2e1}) has the event horizon at $r=r_0$ and hence has the
associated thermodynamics. A standard calculation gives us the ADM mass $M$,
Hawking temperature $T$ and entropy $S$ of the black configuration:
\begin{eqnarray}
&& M=\frac{(8-p)\Omega_{8-p}V_p r_0^{7-p}}{2\kappa^2g^2}\left(1
     +\frac{7-p}{8-p}\sinh^2\alpha \right), \nonumber \\
&& T=\frac{7-p}{4\pi r_0\cosh\alpha}, \nonumber \\
\label{2e10}
&& S=\frac{4\pi \Omega_{8-p}V_p}{2\kappa^2g^2}r_0^{8-p}\cosh\alpha.
\end{eqnarray}
It is somewhat surprising that these thermodynamic quantities are completely
the same as those for the D$p$-branes without $B$ field, just as noticed in
\cite{Mald1,Ali,Bar,CO,Harmark}. The conclusion remains valid even if the
angular rotation is introduced \cite{Harmark}. Here we focus on another
aspect of the thermodynamics of this black configuration: These
thermodynamic quantities are all independent of the parameter $\theta$.
As pointed out above, the parameter $\theta$ characterizes the interpolation
of the solution between the black D$p$-brane without $B$ field and the black
D$(p-2)$-brane with two smeared coordinates and zero $B$ field (a
nonvanishing constant $B$ field along the directions transverse to the
D$(p-2)$-branes can be gauged away). Thus these thermodynamic quantities are
also those of the black D$(p-2)$-branes with two smeared coordinates and no
$B$ field. Therefore there is the thermodynamic equivalence not only between
black D$p$-branes with $B$ field and those without $B$ field, but also
between the black D$p$-branes with $B$ field and black D$(p-2)$-branes with
two smeared coordinates and no $B$ field. This fact is important in the
following discussions.

The thermodynamic quantities (\ref{2e10}) with the charges (\ref{2e3}) and
(\ref{2e5}) satisfy the first law of black hole thermodynamics as expected:
\begin{eqnarray}
\label{first}
dM &=& TdS + \mu_pdq_p +\mu_{p-2}dq_{p-2} \nonumber \\
   &=& TdS +\mu_pV_pT_pdN_p +\mu_{p-2}V_{p-2}T_{p-2}dN_{p-2},
\end{eqnarray}
where $\mu_p$ and $\mu_{p-2}$ are the chemical potentials corresponding
to the total charges $q_p= Q_pV_p$ and $q_{p-2}=V_{p-2}Q_{p-2}$, respectively,
\begin{equation}
\mu_p= \cos\theta \tanh\alpha/g, \ \ \ \mu_{p-2}=\sin\theta \tanh \alpha/g.
\end{equation}
In addition, we notice that in the extremal limit (by taking $r_0
\rightarrow 0$ and $\alpha \rightarrow \infty$, but keeping
$\tilde{R}^{7-p}$ constant),
\begin{equation}
M^2_{\rm ext.}= q_p^2 +q_{p-2}^2,
\end{equation}
which indicates that the bound state (D$(p-2)$, D$p$) is a nonthreshold one.

Now we turn to the field theory limit (decoupling limit) of the bound state
solution (\ref{2e1}), in which the gravity decouples from the field theory
on the worldvolume of D$p$-branes. Following \cite{Mald1,Ali,Lu},
in the decoupling limit
\begin{eqnarray}
\alpha' \rightarrow 0:&& \tan\theta =\frac{\tilde{b}}{\alpha'},
 \ \ \ r=\alpha' u, \ \ \ r_0=\alpha'u_0, \nonumber \\
\label{decoupling}
&& g=\tilde{g}\alpha'^{(5-p)/2},\ \ \ x_{0,1,\cdots,p-2}
  =\tilde{x}_{0,1,\cdots,p-2},\ \ \
     x_{p-1,p}= \frac{\alpha'}{\tilde{b}}\tilde{x}_{p-1,p},
\end{eqnarray}
and $\tilde{b}$, $\tilde{g}$, $u$, $u_0$, and $\tilde{x}_{\mu}$
held fixed, one has the following decoupling limit solution:
\begin{eqnarray}
&& ds^2=\alpha'\left [\left(\frac{u}{R}\right)^{(7-p)/2}\left (-\tilde{f}
    dt^2 +d\tilde{x}_1^2 +\cdots +d\tilde{x}_{p-2}^2
    +\tilde{h}(d\tilde{x}_{p-1}^2 +d\tilde{x}_p^2)\right)
     \right. \nonumber \\
&&~~~~~~~~\left. +\left(\frac{R}{u}\right)^{(7-p)/2}
       \left(\tilde{f}^{-1}du^2
  +u^2 d\Omega^2_{8-p}\right)\right], \nonumber\\
\label{solution}
&& e^{2\phi} = \tilde{g}^2\tilde{b}^2 \tilde{h}\left(\frac{R}{u}
  \right)^{(7-p)(3-p)/2}, \ \ \ B_{p-1,p}
 = \frac{\alpha'}{\tilde{b}}\frac{(au)^{7-p}}{1+(au)^{7-p}},
\end{eqnarray}
where the corresponding RR fields are not exposed explicitly,
\begin{equation}
\tilde{f} = 1 - \left(\frac{u_0}{u}\right)^{7-p},\ \ \tilde{h} = \frac{1}
   {1+(au)^{7-p}}, \ \ \ a^{7-p} = \tilde{b}^2/R^{7-p},
\label{fha}
\end{equation}
and
\begin{equation}
\label{R}
R^{7-p} = \frac{1}{2}(2\pi)^{6-p}\pi^{-(7-p)/2}\Gamma[(7-p)/2]\tilde{g}
  \tilde{b} N_p.
\end{equation}
According to the generalized AdS/CFT correspondence, the solution
(\ref{solution}) is  the dual gravity description
of the noncommutative gauge field theory with gauge group $U(N_p)$ in
($p+1)$ dimensions~\cite{Hashimoto1,Mald1,Ali,Bar,Harmark}. When $a=0$, the
solution (\ref{solution}) reduces to the usual decoupling limit solution of
D$p$-branes without $B$ field \cite{Itzhaki}. This implies that the
noncommutativity effect is weak in field theories for $au<<1$ or at long
distance.

In the decoupling limit, the thermal excitations above the extremality have
the energy $E$, temperature $T$ and entropy $S$:
\begin{eqnarray}
&& E = \frac{(9-p)\Omega_{8-p}\tilde{V}_p}{2(2\pi)^7 (\tilde{g}\tilde{b})^2}
  u_0^{7-p}, \nonumber\\
&& T = \frac{7-p}{4\pi}R^{-\frac{7-p}{2}}u_0^{\frac{5-p}{2}}, \nonumber \\
\label{thermo}
&& S = \frac{2\Omega_{8-p}\tilde{V}_p}{(2\pi)^6(\tilde{g}\tilde{b})^2}
  R^{\frac{7-p}{2}}u_0^{(9-p)/2}.
\end{eqnarray}
Here $\tilde{V}_p=V_{p-2}\tilde{V_2}$ is the spatial volume of the D$p$-brane
after taking the decoupling limit (\ref{decoupling}), and $\tilde{V}_2
= V_2 \tilde{b}^2/\alpha'^2$ is the area of the torus.
Using (\ref{thermo}), one finds the free energy, defined as $F=E-TS$, of the
thermal excitations:
\begin{eqnarray}
F &=& - \frac{(5-p)\Omega_{8-p}\tilde{V}_p}{2(2\pi)^7(\tilde{g}\tilde{b})^2}
 u_0^{7-p} \nonumber \\
\label{freeenergy}
 &=& -\frac{\Omega_{8-p}V_{p-2}\tilde{V}_2}{(2\pi)^7 \tilde{g}^2
  \tilde{b}^2} \frac{5-p}{2}\left(\frac{4\pi}{7-p}\right)^{\frac{2(7-p)}
    {5-p}} R^{\frac{(7-p)^2}{5-p}}T^{\frac{2(7-p)}{5-p}},
\end{eqnarray}
in terms of the temperature. In the decoupling limit, the numbers of two
kinds of branes are constants. Hence the first law of thermodynamics becomes
\begin{equation}
dE = TdS, \ \ \ {\rm and}\ \ \ dF = - SdT.
\end{equation}

In the spirit of the generalized AdS/CFT correspondence, the thermodynamics
of the thermal excitations should be equivalent to that of the corresponding
noncommutative gauge fields at finite temperature $T$ in the large $N$
and strong 't Hooft coupling limit. We notice that these thermodynamic
quantities, after rescaling the string coupling constant as
$\tilde{g}\tilde{b}=\hat{g}$, are exactly the same as those of the black
D$p$-branes without $B$ field in the decoupling limit. This means that in
this supergravity approximation, the thermodynamics of the large $N$
noncommutative and ordinary gauge field theories both in $(p+1)$ dimensions
are equivalent to each other. This also
implies that in the planar limit, the number of the degrees of freedom in
the noncommutative gauge theories coincides with that in the ordinary field
theories not only in the weak coupling limit~\cite{BS}, but also in the
strong coupling limit~\cite{Mald1,Ali,Bar,Harmark}.

Now let us recall the fact that the thermodynamics (\ref{2e10}) of the
black D$p$-branes with nonzero $B$ field is the same as that of the
black D$(p-2)$-branes with two smeared coordinates and zero
$B$ field. In the decoupling limit (\ref{decoupling}), the solution
(\ref{solution}) is described by the quantities of D$p$-branes. For
instance, $R^{7-p}$ is proportional to the number $N_p$ of the coinciding
D$p$-branes [see (\ref{R})]. In fact the ``radius'' $R$ can also be
expressed by quantities of D$(p-2)$-branes. Using (\ref{2e7}), we obtain
\begin{equation}
\label{R2}
R^{7-p} = \frac{1}{2}(2\pi)^{6-p}\pi^{-(7-p)/2}\Gamma[(7-p)/2]
    \tilde{g}\tilde{b}N_{p-2}\times \frac{(2\pi)^2\tilde{b}}{\tilde{V_2}}.
\end{equation}
In the decoupling limit, we are thus led to the relation between the numbers
of D$p$- and D$(p-2)$-branes:
\begin{equation}
\label{number}
\tan\theta = \frac{\tilde{b}}{\alpha'} = \frac{(2\pi)^2\tilde{b}^2}
  {\alpha'\tilde{V}_2}\frac{N_{p-2}}{N_p}\ \ \ \
 \Longrightarrow \ \ \frac{N_{p-2}}{N_p}
 = \frac{\tilde{V}_2}{(2\pi)^2\tilde{b}}.
\end{equation}
Because $\tilde{V}_2$ and $\tilde{b}$ can be kept finite, we can thus
conclude that in the decoupling limit of the D$p$-branes with NS $B$ field,
the number of the D$(p-2)$-branes can be kept finite. This looks different
from the claim by Lu and Roy~\cite{Lu} where they conclude that the number
of the D$(p-2)$-branes becomes infinity in the decoupling limit. This is
so because they take a little different decoupling limit and there
$\tilde{x}_{p-1}$ and $\tilde{x}_p$ are infinitely extended. Mathematically
our decoupling limit becomes the same as theirs by taking
$\tilde{V}_2 \rightarrow \infty$ but keeping $N_{p-2}/\tilde{V}_2 =
N_p/(2\pi)^2\tilde{b}$ finite.

In the decoupling limit (\ref{decoupling}), $\tan\theta \to \infty$ as
$\alpha' \to 0$. If we set $\theta = \pi/2$, the solution~(\ref{2e1})
reduces to the black D$(p-2)$-brane with two smeared coordinates and zero
$B$ field after gauging away the constant value. This shift of the constant
$B$ is allowed in the large $N$ limit~\cite{BS}.\footnote{Alternatively, if
one simply takes the usual D$(p-2)$-brane solution, there is no $B$ field.}
The decoupling limit solution (\ref{solution}) for the black D$p$-brane
with NS $B$ field is thus expected to be related with the solution of black
D$(p-2)$-brane with two smeared coordinates and no $B$ field in the same
decoupling limit. For our convenience, we rewrite the black D$(p-2)$-brane
with two smeared coordinates:
\begin{eqnarray}
&& ds^2 =H^{-1/2}[-fdt^2 +dx_1^2 +\cdots +dx_{p-2}^2 +H (dx_{p-1}^2
 +dx_p^2)] +H^{1/2}(f^{-1}dr^2 +r^2d\Omega_{8-p}), \nonumber\\
&& e^{2\phi}=g^2H^{(5-p)/2}, \ \ \ \
 A^{p-2}_{01\cdots (p-2)} =g^{-1}(H^{-1}-1)\coth\alpha ,\ \ \ \
 B_{p-1,p}=0,
\end{eqnarray}
where $H$ and $f$ are the same as those in (\ref{2e2}). Note that here
$x_{p-1}$ and $x_p$ are two smeared transverse coordinates for the
D$(p-2)$-branes. In the decoupling limit (\ref{decoupling}), we reach
\begin{eqnarray}
&& ds^2=\alpha'\left [\left(\frac{u}{R}\right)^{(7-p)/2}\left (-\tilde {f}
    dt^2 + d\tilde{x}_1^2 +\cdots +d\tilde{x}_{p-2}^2
    + \frac{1}{(au)^{7-p}}(d\tilde{x}_{p-1}^2 + d\tilde{x}_p^2)\right)
     \right. \nonumber \\
&&~~~~~~~~\left. + \left(\frac{R}{u}\right)^{(7-p)/2}
       \left(\tilde{f}^{-1}du^2
   + u^2 d\Omega^2_{8-p}\right)\right], \nonumber\\
\label{dp2}
&& e^{2\phi} = \tilde{g}^2 \tilde{b}^{5-p}(au)^{(7-p)(p-5)/2},
\ \ \ B_{p-1,p}=0,
\end{eqnarray}
where $\tilde{f}$ and $R^{7-p}$ are given in (\ref{fha}) and (\ref{R2}),
respectively. When $\tilde{f}=1$, the solution reduces to that given
in \cite{Lu}. Obviously for $au >>1$, the decoupling solution (\ref{solution})
of the D$p$-brane with NS $B$ field is indeed equivalent to the decoupling
limit solution (\ref{dp2}) of the black D$(p-2)$-branes with two smeared
coordinates and no NS $B$ field, as noticed in \cite{Lu}. (We will also
discuss the equivalence from the thermodynamics point of view shortly.)
Note that the coordinate $u$ corresponds to an energy scale of worldvolume
gauge field theories, and in (\ref{solution}) $au$ reflects the noncommutative
effect of gauge fields. It has been
found that in order for the dual gravity description (\ref{solution})
of noncommutative gauge fields to be valid, $au>>1$ should be satisfied
\cite{Lu,Ali}, in which case $N_p$ can be small and the noncommutativity
effect is strong in the corresponding field theories.

We know that the solution (\ref{solution}) is a dual gravity description of
a noncommutative super Yang-Mills theory with gauge group $U(N_p)$ in ($p+1)$
dimensions. What is the field theory corresponding to the supergravity
solution (\ref{dp2}) for large $au$? To see this, let us note that the
supergravity description~(\ref{dp2}) breaks down for large $au$ since the
effective size of the torus shrinks. Nevertheless, we can make a T-duality
along the directions $\tilde{x}_{p-1}$ and $\tilde{x}_p$. We then obtain
a usual decoupling limit solution of $N_{p-2}$ coincident D$p$-branes
without $B$ field~\cite{Itzhaki}:
\begin{eqnarray}
ds^2 &=&\alpha' \left [\left (\frac{u}{R}\right)^{(7-p)/2}\left(-\tilde{f}dt^2
 +d\tilde{x}_1^2 +\cdots +d\tilde{x}_{p-2}^2 +dx_{p-1}^2 +dx_p^2\right)
    \right.  \nonumber \\
 && ~~~  \left. +\left(\frac{R}{u}\right)^{(7-p)/2}\left(\tilde{f}^{-1}du^2 +
   u^2d\Omega^2_{8-p}\right) \right], \nonumber \\
\label{dp}
 e^{2\phi} &=&  \frac{(2\pi)^4\tilde{g}^2 \tilde{b}^4}{\tilde{V}_2^2}
     \left(\frac{u}{R}\right)^{(7-p)(p-3)/2}, \ \ \ \tilde{B}_{p-1,p}=0.
\end{eqnarray}
This solution describes a ($p+1$)-dimensional ordinary super Yang-Mills theory
with gauge group $U(N_{p-2})$ on the dual torus with area $\hat{V}_2=
(2\pi)^4\tilde{b}^2/\tilde{V}_2$. Since the dual torus is characterized by
the periodicity $x_{p-1,p} \sim x_{p-1,p} + \sqrt{\hat{V}_2}$, the
radii of the dual torus go to zero for $\tilde{V}_2 \to \infty$ and the
($p+1$)-dimensional ordinary super Yang-Mills theory then reduces to a
($p-1$)-dimensional one. This means that if the torus in (\ref{dp2}) is
very large ($\tilde{V}_2 \to \infty$), the solution is a dual
gravity description of a ($p-1)$-dimensional ordinary super Yang-Mills
theory with gauge group $U(\infty)$. The ($p-1$)-dimensional theory has the
Yang-Mills coupling constant
\begin{equation}
\label{const1}
g^2_{\rm YM}=(2\pi)^{p-4}\tilde{g},
\end{equation}
while the coupling constant of the $(p+1)$-dimensional noncommutative gauge
field is $g^2_{\rm YM}=(2\pi)^{p-2}\tilde{g}\tilde{b}$. Thus we reach the
equivalence argued by Lu and Roy \cite{Lu} between the noncommutative
super Yang-Mills theory in ($p+1)$ dimensions and the ordinary one
with gauge group $U(\infty)$ in ($p-1$) dimensions.

This equivalence can also be understood from a T-duality of the decoupling
limit solution~(\ref{solution}) for the D$p$-branes with $B$ field.
A usual T-duality transformation~\cite{BHO} is, however, not enough for this
purpose since the resulting radius for the torus is not large after the usual
T-duality in the presence of $B$ field. This can be remedied if we use more
general T-duality transformation $SL(2,{\bf Z})$~\cite{GPR} as described
in refs.~\cite{Douglas,Hashimoto2}. The duality transformation
\begin{equation}
\rho \to \frac{a\rho+b}{c\rho+d}, \;\;
\rho \equiv \frac{{\tilde V}_2}{(2\pi)^2 \alpha'}\left( B_{p-1,p}
 +i \sqrt{G_{(p-1)(p-1)}G_{pp}}\right),
\end{equation}
gives a dual solution
\begin{eqnarray}
&& ds^2 = \alpha' \left[ \left(\frac{u}{R}\right)^{(7-p)/2}
 \left( -\tilde{f} dt^2 + d\tilde{x}_1^2 +\cdots +d\tilde{x}^2_{p-2}+
 dx_{p-1}^2  +dx_p^2 \right) \right. \nonumber \\
 &&~~~~~~ \left.
 +\left(\frac{R}{u}\right)^{(7-p)/2}\left({\tilde f}^{-1}
 du^2 +u^2 d\Omega_{8-p}^2 \right) \right], \nonumber \\
\label{morita}
&& e^{2\phi} = \frac{(2\pi)^4\tilde{g}^2 \tilde{b}^4}{\tilde{V}_2^2}
     \left(\frac{u}{R}\right)^{(7-p)(p-3)/2}, \ \ \ \tilde{B}_{p-1,p}=
    \frac{\alpha'}{\tilde{b}},
\end{eqnarray}
by choosing $c=-1$ and $d=\tilde{V}_2/(2\pi)^2 \tilde{b}$ when the latter is
an integer. Note that $d=N_{p-2}/N_p$ must be a rational number. If this is
not an integer, after some steps of Morita equivalence transformation
following~\cite{Hashimoto2}, one can reach a solution like (\ref{morita}).
For $p=3$ and $\tilde{f}=1$, the solution~(\ref{morita}) reduces to the case
discussed in \cite{Douglas,Hashimoto2}. Note that the solution (\ref{morita})
is the same as (\ref{dp}) except that the former has a nonvanishing constant
$B$ field while the latter has zero $B$ field. The solution (\ref{morita})
describes a twisted ordinary super Yang-Mills theory with gauge group
$U(N_{p-2})$ in $(p+1)$ dimensions, living on the dual torus with area
$\hat{V}_2 = (2\pi)^4\tilde{b}^2/\tilde{V}_2$. For $\tilde{V}_2 \to \infty$,
however, the theory reduces to a ($p-1$)-dimensional ordinary super Yang-Mills
theory. Therefore we again arrive at the conclusion that the
($p+1$)-dimensional noncommutative gauge field is equivalent to an ordinary
gauge field with gauge group $U(\infty)$ in ($p-1$) dimensions for
$\tilde{V}_2 \to \infty$, from the point of view of dual gravity description.

Next let us address another evidence to render support of the above
equivalence from the viewpoint of thermodynamics. We find that the
thermodynamics of decoupling limit solution (\ref{dp2}) of the D$(p-2)$-branes
is completely the same as those in (\ref{thermo}). The worldvolume theory of
the (D$(p-2)$, D$p$) bound states (\ref{2e1}) (or $N_p$ coinciding D$p$-branes
with $B$ field) is a noncommutative gauge field theory with gauge group
$U(N_p)$ in $(p+1)$ dimensions with two dimensions compactified on a torus,
while the worldvolume theory is an ordinary gauge field theory with the
same gauge group $U(N_p)$ if the NS $B$ field is absent. The above
equivalence of the descriptions of the (D$(p-2)$, D$p$) bound states
implies that the bound states can also be described by ordinary gauge field
theories in $(p+1)$ dimensions. Moreover, the equivalence is valid also
between the $(p+1)$-dimensional noncommutative $U(N_p)$ theory and the
$(p-1)$-dimensional $U(\infty)$ ordinary theory in the limit ${\tilde V}_2
\to \infty$ for the reason described above. For finite volume, the
equivalence is between the $(p+1)$-dimensional noncommutative $U(N_p)$ gauge
field and a (twisted) ordinary $U(N_{p-2})$ gauge field with the
relation~(\ref{number}), the latter living on a dual torus. In this case,
the Yang-Mills coupling constant is
\begin{equation}
\label{const2}
g^2_{\rm YM}=\frac{(2\pi)^p \tilde{g}\tilde{b}^2}{\tilde{V}_2},
\end{equation}
for the $(p+1)$-dimensional ordinary gauge field theory. As a self-consistency
check, one may find that the coupling constant (\ref{const1}) can also be
obtained from (\ref{const2}) after a trivial dimensional reduction. In the
following sections we will further discuss the equivalence of the descriptions
and the relation (\ref{number}) from the point of view of probe branes.

\sect{The static probes: thermodynamics}

The D-brane probe is a useful tool to explore the structure of D-brane
bound states (see for example \cite{Dou}-\cite{Youm} and references
therein). Recently the D-brane probes have been used to check a certain
aspect of AdS/CFT correspondence~\cite{AA}-\cite{Cai}. In this section
we consider the static interaction potentials (thermodynamics) of two
kinds of probes in the background of (D$(p-2)$, D$p$) bound states.
One of them is a bound state probe consisting of D$(p-2)$- and D$p$-branes,
or D$p$-brane probe with $B$ fields, or noncommutative D$p$-brane probe;
the other is a D$(p-2)$-brane probe.
Let us first discuss the noncommutative D$p$-brane probe.

\subsection{A noncommutative D$p$-brane probe}

Due to the presence of the nonvanishing NS $B$ field in the noncommutative
D$p$-brane probe, the probe should have the following action:
\begin{equation}
\label{3e1}
S_p= -T_p\int d^{p+1}x e^{-\phi}\sqrt{-\det(G_{ab} +{\cal F}_{ab})}
  +T_p\int A^{p} +T_p \int A^{p-2}\wedge {\cal F},
\end{equation}
where ${\cal F}_{ab} = (2\pi \alpha')F_{ab} +B_{ab}$, $F_{ab}$ is
the gauge field strength on the worldvolume of the D$p$-brane and $B_{ab}$
is the NS $B$ field. Here we set $F_{ab}=0$. As demonstrated in (\ref{2e1}),
the occurrence of the NS $B$ field in the D$p$-branes is always accompanied
by the appearance of D$(p-2)$-branes in the system, forming (D$(p-2)$, D$p$)
bound states. The probe can be regarded as a bound state probe consisting
of D$(p-2)$- and D$p$-branes for the following reasons:
(1) The probe has the tension $T_p\sqrt{1+\tan^2\theta}$, the same as the
(D$(p-2)$, D$p$) bound states;
(2) We will see shortly that the static interaction potential of the probe
vanishes in the background of the nonthreshold (D$(p-2)$, D$p$) bound states;
(3) Except for the source $A^{p}$ of the D$p$-branes, the source $A^{p-2}$ of
the D$(p-2)$-branes also occurs in the action (\ref{3e1});
(4) From its thermodynamics we will obtain further evidence of this
interpretation. Because of the presence of the NS $B$ field,
we can also view the probe as a noncommutative D$p$-brane probe.

Substituting the solution (\ref{2e1}) into the probe action (\ref{3e1}),
one has
\begin{equation}
\label{3e2}
S_p=-\frac{T_pV_p}{g\cos\theta}\int d\tau H^{-1}[\sqrt{f}-1+H_0-H],
\end{equation}
where we have subtracted a constant potential at spatial infinity and
\begin{equation}
H_0= 1+\left(\frac{\tilde{R}}{r}\right)^{7-p}=1+\frac{r_0^{7-p}
   \sinh\alpha\cosh\alpha}{r^{7-p}}.
\end{equation}
In the extremal limit ($f=1$), the static interaction potential vanishes,
which verifies that the probe is a bound state of D$(p-2)$-branes and
D$p$-branes because the source is nonthreshold bound states of D$(p-2)$-
and D$p$-branes. Unless the probe is such a kind of bound state, the static
potential will no longer vanish. In the non-extremal
background, the interaction potential always exists.

Now suppose the probe is moved from  spatial infinity to the horizon
of the source \cite{KT}. From (\ref{3e2}) we can obtain the potential
difference (which is just the potential at the horizon because we have set
the potential zero at spatial infinity)
\begin{equation}
\label{3e4}
U_p|_{r=r_0}= \frac{T_p V_p}{g\cos\theta}\left (1-\tanh\alpha \right ).
\end{equation}
Since the tension of the probe is $T_p\sqrt{1+\tan^2\theta}$,
it is easy to show that the first term is just the mass of the probe because
\begin{equation}
m_p=\frac{T_pV_p}{g}\sqrt{1+\tan^2\theta}=\frac{T_pV_p}{g\cos\theta}.
\end{equation}
The second term in (\ref{3e4}) has the following interpretation. Let us
denote the numbers of D$p$-branes and D$(p-2)$-branes in the probe by
$\delta N_p$ and $\delta N_{p-2}$, respectively. We then have
\begin{eqnarray}
\mu_p V_pT_p \delta N_p &+& \mu_{p-2}V_{p-2}T_{p-2}\delta N_{p-2}
         \nonumber\\
 &&~~~~~~~~~~= \frac{T_p V_p}{g\cos\theta}\left [\cos^2\theta +
   \sin\theta \cos\theta \frac{V_{p-2}T_{p-2}}{V_pT_p}
    \frac{\delta N_{p-2}}{\delta N_p}\right] \delta N_p, \nonumber \\
\label{n_p}
 &&~~~~~~~~~~= \frac{T_pV_p}{g\cos\theta}\tanh\alpha~\delta N_p,
\end{eqnarray}
where in obtaining the third line we have used the fact that the form of the
tension of the probe implies that the number of the branes obey
$\delta N_{p-2}/\delta N_p=\tan\theta V_pT_p/(V_{p-2}T_{p-2})$. When
$\delta N_p=1$, this quantity (\ref{n_p}) gives the second term in
(\ref{3e4}). This process satisfies the first law of thermodynamics
(\ref{first}). In fact eq.~(\ref{3e4}) reduces to (\ref{first}) with $dM=m_p$,
$U_p|_{r=r_0}= TdS$ and (\ref{n_p}). It follows that the potential of the
probe is converted into heat energy and is absorbed by the source when the
probe moves to the horizon from  spatial infinity. Our calculation also
shows that the probe is a bound state of D$(p-2)$- and D$p$-branes.

Now we consider the decoupling limit of the static probe action. In this
limit, we obtain
\begin{equation}
\label{3e7}
S_p=-\frac{V_{p-2}\tilde{V}_2}{(2\pi)^p\tilde{g}\tilde{b}}\int d\tau
  \left(\frac{u}{R}\right)^{7-p}\left[\sqrt{\tilde{f}}-1+
    \frac{u_0^{7-p}}{2u^{7-p}}\right].
\end{equation}
{}From the action we can also obtain the free energy of the probe at the
temperature $T$, which is just the Hawking temperature of the source given
in (\ref{thermo}):
\begin{equation}
\label{3e8}
F_p=\frac{V_{p-2}\tilde{V}_2}{(2\pi)^p\tilde{g}\tilde{b}}
  \left(\frac{u}{R}\right)^{7-p}\left[\sqrt{\tilde{f}}-1+
    \frac{u_0^{7-p}}{2u^{7-p}}\right].
\end{equation}

In the generalized AdS/CFT correspondence, the thermodynamics (\ref{thermo})
is equivalent to that of noncommutative supersymmetric gauge fields
with gauge group $U(N_p)$ in the large $N$ and strong coupling limit (within
the valid regime of dual gravity description). From the viewpoint of field
theory, the thermodynamics corresponds to that of the gauge field in the
Higgs branch, where the gauge group is not broken and hence the vacuum
expectation values of scalars vanish. According to the interpretation of
a D-brane probe action \cite{Mald3}, the thermodynamics of a D-brane probe
can be regarded as the thermodynamics of the supersymmetric gauge field
in the Coulomb branch (Higgs phase) \cite{AA,Kirit}, in which the original
gauge group is broken; some vacuum expectation values of scalar fields
do not vanish; and the distance $u$ between the probe and the source can
be viewed as a energy scale in the gauge fields. This interpretation of
thermodynamics of D-brane probe turns out to be consistent with
the expectation on the field theory side \cite{AA,Kirit}.

Since the rescaled string coupling is $\hat{g}=\tilde{g}\tilde{b}$,
we find from (\ref{3e8}) that in the decoupling limit, the free energy
of a noncommutative D$p$-brane probe in the noncommutative D$p$-brane
background ((D$(p-2)$, D$p$) bound states) is exactly the same as that of an
ordinary D$p$-brane probe in the D$p$-brane background without NS $B$ field
(for the interaction potential of the latter see \cite{Kirit}). Thus,
in the supergravity approximation, the thermodynamics of noncommutative
gauge fields remains the same as the ordinary case both in the Higgs and
Coulomb branches. We thus conclude that in the large $N$ limit, the number
of the degrees of freedom of noncommutative gauge fields coincides with
the ordinary case, not only in the weak coupling limit, but also in the
strong coupling limit.

Now we consider the difference between the interaction potentials (free
energy) of the probe at the infinity and at the horizon in the decoupling
limit, and compare it with the asymptotically flat case already discussed
before. Note that the free energy still vanishes at the infinity ($u \to
\infty$) as can be seen in (\ref{3e8}). Thus the difference in the free
energies is just the free energy of the probe at the horizon $u_0$:
\begin{eqnarray}
F_p|_{u=u_0} &=&- \frac{V_{p-2}\tilde{V}_2}{2(2\pi)^p\tilde{g}\tilde{b}}
    \left( \frac{u_0}{R}\right)^{7-p} \nonumber \\
\label{3e9}
    &=&  - \frac{V_{p-2}\tilde{V}_2}{2(2\pi)^{p}\tilde{g}\tilde{b}}
   \left(\frac{4\pi RT}{7-p}\right)^{\frac{2(7-p)}{5-p}}.
\end{eqnarray}
We find that the free energy of the probe at the horizon (\ref{3e9}) has
the relation with that of the source (\ref{freeenergy}) as
\begin{equation}
F_p|_{u=u_0}=\frac{dF}{dN_p}\delta N_p,
\label{fe}
\end{equation}
with $\delta N_p=1$. Note that we are considering a D$p$-brane with $B$ field
in the background of $N_p$ D$p$-branes. Therefore in the large $N_p$ limit
(that is $N_p>>1$), the probe free energy is expected to be
\begin{equation}
F_p|_{u=u_0} \approx F(N_p+1)-F(N_p),
\end{equation}
consistent with (\ref{fe}). This relation supports the argument that the
non-extremal D$p$-branes is {\it located} at the horizon. (In the next
section we will further show that indeed D$p$-branes can be located at the
horizon). This also supports the interpretation of thermodynamics of probe
branes given in \cite{AA,Kirit}, because the probe brane at the horizon of
the source can be considered to coincide with source branes and the gauge
symmetry is restored and the probe brane can be seen as a part of
the source branes in this case.

\subsection{A D$(p-2)$-brane probe}

Let us next consider a D$(p-2)$-brane probe. The action of a D$(p-2)$-brane
in the background of the (D$(p-2)$, D$p$) bound states (\ref{2e1}) is
\begin{equation}
\label{dp-2action}
S_{p-2}=-T_{p-2}\int d^{p-1}x e^{-\phi}\sqrt{-\det G_{ab}}+T_{p-2}
\int A^{p-2}.
\end{equation}
Substituting the background solution (\ref{2e1}) into the action yields,
for a static probe,
\begin{equation}
\label{3e13}
S_{p-2}=-\frac{T_{p-2}V_{p-2}}{g}\int d\tau H^{-1}\left [H^{1/2}h^{-1/2}
\sqrt{f} -(1-H_0)\sin\theta -H \right ],
\end{equation}
where we have also subtracted a constant potential so that the interaction
potential vanishes at spatial infinity. Note that the static interaction
potential is quite different from that of the noncommutative D$p$-brane probe
in the same background (\ref{2e1}). Indeed the potential (\ref{3e13})
does not vanish even when the background is sent to the extremal limit of the
solution. This is consistent with the fact that the source is a nonthreshold
bound state consisting of D$(p-2)$- and D$p$-branes.

As in the noncommutative D$p$-brane probe, let us first consider the
asymptotically flat background. In this case, we find that the deference
between the potentials at the spatial infinity and at the horizon is
\begin{equation}
\label{3e14}
U_{p-2}|_{r=r_0} =\frac{V_{p-2}T_{p-2}}{g}\left (1
       -\sin\theta \tanh\alpha \right).
\end{equation}
The first term is the mass of the probe $m_{p-2}=T_{p-2}V_{p-2}/g$, while
the second term is equal to $\mu_{p-2}V_{p-2}T_{p-2}\delta N_{p-2}$
with $\delta N_{p-2}=1$ since we are considering a probe D$(p-2)$-brane.
Therefore the D$(p-2)$-brane probe falling to the horizon from the spatial
infinity satisfies the first law (\ref{first}) again with $dM=m_{p-2}$,
$\delta N_{p-2}=1$, and $\delta N_p=0$. The potential difference of the
D$(p-2)$-brane probe is converted into heat energy at the horizon and
thereby is absorbed by the source.

Comparing (\ref{3e2}) and (\ref{3e13}), {\it a priori} one may think that
they are quite different and there seems to be no relation between them.
Actually once the decoupling limit (\ref{decoupling}) is taken, one may find
that there is a close relation between (\ref{3e2}) and (\ref{3e13}). In the
decoupling limit, the action of the D$(p-2)$-brane probe becomes
\begin{equation}
S_{p-2}=-\frac{V_{p-2}}{(2\pi)^{p-2}\tilde{g}}\int d\tau
      \left (\frac{u}{R}\right)^{7-p}\left [
    \sqrt{\frac{1+(au)^{7-p}}{(au)^{7-p}}}\sqrt{\tilde{f}}-1
      +\frac{u_0^{7-p}}{2u^{7-p}}\right ].
\end{equation}
As mentioned above, the validity of the dual gravity description of gauge
field theories requires $au>>1$. The above action then reduces to
\begin{equation}
S_{p-2}= -\frac{V_{p-2}}{(2\pi)^{p-2}\tilde{g}}\int d\tau
      \left (\frac{u}{R}\right)^{7-p}\left [
    \sqrt{ \tilde{f}}-1
      +\frac{u_0^{7-p}}{2u^{7-p}}\right ].
\end{equation}
The corresponding free energy of the probe at the distance $u$ is
\begin{equation}
F_{p-2}= -\frac{V_{p-2}}{(2\pi)^{p-2}\tilde{g}}
      \left (\frac{u}{R}\right)^{7-p}\left [
    \sqrt{ \tilde{f}}-1
      +\frac{u_0^{7-p}}{2u^{7-p}}\right ].
\end{equation}
At the horizon $u_0$ it is
\begin{eqnarray}
F_{p-2}|_{u=u_0} &=&-\frac{V_{p-2}}{2(2\pi)^{p-2}\tilde{g}}
  \left(\frac{u_0}{R} \right)^{7-p}  \nonumber \\
\label{3e17}
    &=& -\frac{V_{p-2}}{2(2\pi)^{p-2}\tilde{g}}
     \left (\frac{4\pi RT}{7-p}\right)^{\frac{2(7-p)}{5-p}}.
\end{eqnarray}
Comparing the free energy (\ref{3e17}) with the one (\ref{3e9}) of the
noncommutative D$p$-brane probe, one may find that they are the same up to a
different prefactor. Consequently the free energy of $\delta N_{p-2}$
D$(p-2)$-branes is the same as that of $\delta N_p$ noncommutative D$p$-branes
if the relation
\begin{equation}
\label{3e18}
\frac{\delta N_{p-2}}{\delta N_p}= \frac{\tilde{V}_2}{(2\pi)^2 \tilde{b}},
\end{equation}
is obeyed. We see that this relation coincides with eq.~(\ref{number}). Note
that the relation (\ref{number}) is derived from the two equivalent
descriptions of the bound state source, while (\ref{3e18}) is obtained from
the equivalence of probes in the same background. In other words, the probe
consisting of $\delta N_p$ noncommutative D$p$-branes is equivalent to the
probe consisting of $\delta N_{p-2}$ D$(p-2)$-branes since they get the same
response in the same background. Furthermore, we find
\begin{equation}
F_{p-2}|_{u=u_0}=\frac{dF}{dN_{p-2}}\delta N_{p-2},
\end{equation}
with $\delta N_{p-2}=1$. When $N_{p-2}>>1$, once again, we have
\begin{equation}
F_{p-2}|_{u=u_0}\approx F(N_{p-2}+1)-F(N_{p-2}).
\end{equation}
This implies that from the point of view of the D$(p-2)$-brane probe,
the bound state source (D$(p-2)$, D$p$) can be viewed as
$N_{p-2}$ coincident D$(p-2)$-branes with two smeared coordinates and zero
NS $B$ field, while from the noncommutative D$p$-brane probe, the
source is $N_p$ coincident D$p$-branes with nonzero NS $B$ field. Therefore
from the thermodynamics of probe branes, we again find that the
bound states (D$(p-2)$, D$p$) have two equivalent descriptions.

\sect{The dynamical probes: absorbing or scattering}

In this section we will consider the dynamical aspect of the two kinds of
probes discussed in the previous section.

\subsection{The noncommutative D$p$-brane probe}

To investigate the dynamics of the probe, it is convenient to take static
gauge: $\tau =t$, $x_i$ act just as the worldvolume coordinates and other
transverse coordinates depend only on $\tau$. In the background (\ref{2e1}),
the action (\ref{3e1}) of the noncommutative D$p$-brane probe reduces to
\begin{equation}
\label{4e1}
S_p=-\frac{T_pV_p}{g\cos\theta}
  \int d\tau H^{-1}[\sqrt{f-H(f^{-1}\dot{r}^2
  + r^2 \dot{\Omega}_{8-p}^2)}-1 + H_0 - H],
\end{equation}
where an overdot denotes the derivative with respect to $\tau$. In the
decoupling limit, the action becomes
\begin{eqnarray}
\label{4e2}
S_p =-m_p\int d\tau \left(\frac{u}{R}\right)^{7-p}\left [
 \sqrt{\tilde{f} -\left(\frac{R}{u}\right)^{7-p}\left(\tilde{f}^{-1}
 \dot{u}^2 +u^2 \dot{\Omega}^2_{8-p}\right)}-1
 +\frac{u_0^{7-p}}{2u^{7-p}}\right],
\end{eqnarray}
where $m_p=V_{p-2}\tilde{V}_2/[(2\pi)^p\tilde{g}\tilde{b}]$ is the mass of
the probe. With the definition of the isotropic coordinates
\begin{equation}
\tilde{f}^{-1} du^2 +u^2 d\Omega_{8-p}^2 = u^2 {\rho}^{-2}
     (d{\rho}^2 +{\rho}^2 d\Omega^2_{8-p}),
\end{equation}
where
\begin{equation}
u^{7-p}=\rho^{7-p}\left ( 1+\frac{u_0^{7-p}}{4\rho^{7-p}} \right)^2,
\end{equation}
we can define the velocity of the probe as
\begin{equation}
\tilde{f}^{-1}\dot{u}^2 + u^2 \dot{\Omega}_{8-p}^2
 \equiv u^2{\rho}^{-2}v^2.
\end{equation}
In the low velocity and long distance approximation, expanding (\ref{4e2})
yields
\begin{equation}
S_p =\int d\tau [\frac{1}{2}m_p v^2 -{\cal V}(\rho,v)
 +{\cal O}(1/\rho^{2(7-p)})],
\end{equation}
where the interaction potential ${\cal V}$ is
\begin{equation}
\label{4e7}
{\cal V}(\rho, v)=-m_p\frac{u_0^{7-p}}{\rho^{7-p}}
  \left \{ \frac{9-p}{4(7-p)} v^2
 +\frac{1}{8}\left [\left(\frac{u_0}{R}\right)^{7-p} +
  \left(\frac{R}{u_0}\right)^{7-p}v^4 \right] \right\}
\end{equation}
When $\theta =0$, the background (\ref{2e1}) reduces to the black D$p$-brane
without $B$ and the probe action (\ref{4e1}) to the one for a D$p$-brane
without $B$ field. The interaction potential (\ref{4e7}) therefore is also
the one for a D$p$-brane probe in other D$p$-brane background. Up to order
$v^4$, it can be seen that (\ref{4e7}) reduces to the result in \cite{Mald3}
for a D$p$-brane in other D$p$-brane background. This interaction potential
can be reproduced in the one-loop calculation of the noncommutative field
theory. Using the relation between the phase shift of scattering and the
potential~\cite{Chep1}
\begin{equation}
\delta (\rho,v) =-\int^{\infty}_0 d\tau {\cal V}[\rho(\tau),v],
\ \ \ \ \rho^2(\tau) =\rho^2 +v^2 \tau^2,
\end{equation}
we obtain the phase shift of the probe. Writing ${\cal V}(\rho,v)=
\lambda(v)\rho^{-(7-p)}$, we find
\begin{equation}
\delta (\rho,v) = -\frac{B\left(\frac{1}{2}, \frac{6-p}{2}\right)}
{2v \rho^{6-p}} \lambda (v),
\end{equation}
where $B$ is the beta function.

Next we discuss the classical motion of the noncommutative D$p$-brane probe
near the horizon of the source (that is, in the decoupling limit). For
simplicity, let us consider the case of the probe with angular momentum
only in a single direction (say, $\phi$-direction). Following
refs.~\cite{Liu,Youm}, from (\ref{4e2}) we
obtain its angular momentum
\begin{equation}
L=\frac{m_p u^2 \dot{\phi}}{
 \sqrt{\tilde{f} -\left(\frac{R}{u}\right)^{7-p}
 \left(\tilde{f}^{-1}\dot{u}^2 +u^2\dot{\phi}^2\right)} }.
\end{equation}
The energy of the probe is
\begin{equation}
E = \frac{m_p\left(\frac{u}{R}\right)^{7-p}\tilde{f}}{
 \sqrt{\tilde{f} -\left(\frac{R}{u}\right)^{7-p}
 \left(\tilde{f}^{-1}\dot{u}^2 +u^2\dot{\phi}^2\right)}}
 -m_p\left(\frac{u}{R}\right)^{7-p}\left (1
 -\frac{u_0^{7-p}}{2u^{7-p}} \right ).
\end{equation}
With the relation
\begin{equation}
E= \frac{1}{2}m_p \dot{u}^2 + V(u),
\end{equation}
one may obtain an effective central potential of the radial motion of the
probe
\begin{equation}
\label{4e14}
V(u)=E\left[1 -\frac{m_p \tilde{f}^2}{2E}\left(\frac{u}{R}\right)^{7-p}
 \left (1-\frac{\tilde{f}}{{\cal A}^2}\right)\right]
 +\frac{L^2 \tilde{f}^3}{2m_p u^2 {\cal A}^2}.
\end{equation}
where
\begin{equation}
{\cal A}= 1-\frac{u_0^{7-p}}{2u^{7-p}}+\frac{E}{m_p}
 \left(\frac{R}{u}\right)^{7-p}.
\end{equation}
The qualitative features of the motion of the probe can be understood by
finding out the turning points, at which $\dot{u}=0$.

Let us first discuss
the case of extremal background (or in the background of the nonthreshold
bound state (D$(p-2)$, D$p$). One has $\tilde{f}=1$. From the effective
central potential one can see that if the angular momentum vanishes,
there is no turning point for the probe. It follows that the probe will
be captured by the source. When the angular momentum does not vanish,
the potential (\ref{4e14}) reduces to
\begin{equation}
V(u)=E\left \{1-\frac{1}{2}\left(\frac{u}{u_*}\right)^{7-p}
 \left [1-\frac{1}{\left(1+(u_*/u)^{7-p}\right)^2}\right] \right\}
 +\frac{Eu_{**}^2}{2u^2 \left (1+(u_*/u)^{7-p}\right)^2},
\end{equation}
where we have introduced two characterizing lengths
\begin{equation}
u_*=R\left(\frac{E}{m_p}\right )^{1/(7-p)}, \ \ \
u_{**} = L\left(\frac{1}{m_pE}\right)^{1/2}.
\end{equation}
The turning point satisfies the following equation:
\begin{equation}
2 +\left(\frac{u_*}{u_c}\right)^{7-p} =\left(\frac{u_{**}}{u_c}\right)^2.
\end{equation}
If $u_*/u >>1$, namely, very near the source branes, the turning point
is
\begin{equation}
u_c= \left (\frac{u_*^{7-p}}{u_{**}^2}\right)^{1/(5-p)}.
\end{equation}

In the non-extremal background, there may exist some points satisfying
the turning-point condition $\dot{u}=0$. From the effective potential
(\ref{4e14}) we find that the horizon, where $\tilde{f}=0$, must be one
of those points, regardless of the angular momentum. In particular,
we notice that the central force exerted on the probe, defined as
$F(u)=-dV(u)/du$, vanishes at the horizon. It means that once
the probe reaches the horizon, it can stay at the horizon since the horizon
is the turning point and the central force is zero there. In the previous
section we have shown that the non-extremal branes are ``located''
at the horizon from the point of view of thermodynamics of a probe brane.
Here we provide another evidence to support the argument from the dynamical
aspect of a probe brane.

\subsection{The D$(p-2)$-brane probe}

In this subsection we consider the dynamics of a D$(p-2)$-brane in
the background of (D$(p-2)$, D$p$) bound states in the decoupling limit.
In this case, the worldvolume is $(t, x_1, \cdots, x_{p-2})$, in the
static gauge, one has the action of the probe
\begin{eqnarray}
S_{p-2} &=& - \frac{T_{p-2}V_{p-2}}{g}\int d\tau H^{-1}\left [(Hh^{-1})^{1/2}
 \sqrt{f -h(\dot{x}_{p-1}^2 +\dot{x}_p^2)
 -H(f^{-1}\dot{r}^2 + r^2 \dot{\Omega}_{8-p}^2)} \right. \nonumber \\
&& \hspace{4cm} - \left. (1-H_0)\sin\theta -H\right],
\end{eqnarray}
In the decoupling limit, it reduces to
\begin{eqnarray}
S_{p-2} &=& - \; m_{p-2} \int d\tau \left(\frac{u}{R}\right)^{7-p}
 \left [\sqrt{\frac{1}{(au)^{7-p}\tilde{h} }}
 \sqrt{\tilde{f} -\tilde{h}\left (\dot{\tilde x}_{p-1}^2
 +\dot{\tilde x}_p^2 \right) -\left(\frac{R}{u}\right)^{7-p}
 \left(\tilde{f}^{-1} \dot{u}^2 +u^2\dot{\Omega}_{8-p}^2 \right)}
 \right.  \nonumber \\
\label{4e20}
& & \hspace{4cm} - \left. 1 +\frac{u_0^{7-p}}{2u^{7-p}}\right ].
\end{eqnarray}
Here $m_{p-2}= V_{p-2}/[(2\pi)^{p-2}\tilde{g}]$ is the mass of the probe.
When $au >>1$, this action approximates to
\begin{eqnarray}
S_{p-2} &=& - m_{p-2} \int d\tau \left(\frac{u}{R}\right)^{7-p}
 \left [ \sqrt{\tilde{f} -\frac{1}{(au)^{7-p}}\left (\dot{\tilde x}_{p-1}^2
 +\dot{\tilde x}_p^2 \right) -\left(\frac{R}{u}\right)^{7-p}
 \left(\tilde{f}^{-1} \dot{u}^2 +u^2\dot{\Omega}_{8-p}^2 \right)}
 \right.  \nonumber \\
\label{4e21}
& & \hspace{4cm} - \left. 1 +\frac{u_0^{7-p}}{2u^{7-p}}\right ].
\end{eqnarray}
In fact, this is also the action of a D$(p-2)$-brane probe in the
background produced by the source D$(p-2)$-branes (\ref{dp2}). In the
extremal limit, up to ${\cal O}(v^4)$, its motion is a geodesic
of the following moduli space:
\begin{equation}
ds^2_m= du^2 +u^2 d\Omega_{8-p}^2 +
 \frac{1}{\tilde{b}^2}\left (d\tilde{x}_{p-1}^2+ d\tilde{x}_p^2\right).
\end{equation}
Therefore, in this approximation, the D$(p-2)$-brane probe moves as
in a flat space. Namely, in the large noncommutative effect limit,
the effect of the $N_p$ coincident D$p$-branes in the source on the
motion of a D$(p-2)$-brane probe vanishes. The source looks like
the one consisting of only D$(p-2)$-branes with two smeared coordinates
and no $B$ field.

In the large $au>>1$ limit, if one does not consider the motion of the
probe along the relative transverse directions (that is, setting
$\dot{\tilde x}_{p-1} =\dot{\tilde{x}}_p=0$ in the the action (\ref{4e20})),
the action of the probe then reduces to that (\ref{4e2}) of
a noncommutative D$p$-brane probe, except a difference in the mass of probes.
If the mass of both probes is equal to each other, then they have the same
interaction potential and the same phase shift depending on the distance
and the velocity. Because we are considering only a single D-brane probe,
to match the mass of probes, we should have $\delta N_p m_p=\delta N_{p-2}
m_{p-2}$, from which one gets the relation (\ref{3e18}) again. Thus, from
the point of view of the dynamics of probes, we see again the equivalence
between $\delta N_p$ noncommutative D$p$-branes and $\delta N_{p-2}$
D$(p-2)$-branes with two smeared coordinates and no $B$ field.

Without the motion along the relative transverse directions,
similarly to the case of the D$p$-brane probe, we also obtain an effective
central potential for the D$(p-2)$-brane probe as
\begin{equation}
\label{dp2pot}
V(u)= E\left \{ 1-\frac{m_{p-2}\tilde{f}^2}{2E}
 \left(\frac{u}{R}\right)^{7-p}\left [1-\frac{{\cal C} \tilde{f}}
 {{\cal B}^2}\right] \right\}
 +\frac{L^2 \tilde{f}^3}{2m_{p-2}u^2 {\cal B}^2},
\end{equation}
where we have not taken the large $au$ limit and
\begin{equation}
{\cal C}=\frac{1+(au)^{7-p}}{(au)^{7-p}}, \ \ \
{\cal B}=1-\frac{u_0^{7-p}}{2u^{7-p}}
 +\frac{E}{m_{p-2}}\left(\frac{R}{u}\right)^{7-p}.
\end{equation}
Due to the appearance of ${\cal C}$, the motion of the probe D$(p-2)$-brane
is a little different from that of the D$p$-brane probe. However, we find
that the horizon of the background is still the turning point of the probe
and the central force on the D$(p-2)$-brane probe vanishes. This implies that
the D$(p-2)$-brane probe can also stay at the horizon. It is also consistent
with the result from the analysis of thermodynamics of the probe. Indeed, as
an ingredient of the bound state (D$(p-2)$, D$p$), non-extremal
D$(p-2)$-branes should also be located at the horizon of the background.

\sect{Conclusions}

As is well known by now, the worldvolume coordinates of D$p$-branes will
become noncommutative if a nonvanishing constant NS $B$ field is present
on the worldvolume of the D$p$-branes. The worldvolume theory is then the
super Yang-Mills theory in a noncommutative space (noncommutative
gauge field theory). Each of the nonthreshold (D$(p-2)$, D$p$) bound states
($2 \le p \le 6$) can be viewed as a D$p$-brane bound state with a
nonvanishing NS $B$ field of rank two.

In this paper we have investigated two equivalent descriptions of the
nonthreshold (D$(p-2)$, D$p$) bound states in the dual gravity description.
In the decoupling limit, the bound states can be described as D$p$-branes
with nonvanishing NS $B$ field, and then the worldvolume theory is a
noncommutative gauge field with gauge group $U(N_p)$ in ($p+1$) dimensions
(with two dimensions compactified on a torus in our case) if the number of
the coincident D$p$-branes is $N_p$. On the other hand, the nonthreshold
(D$(p-2)$, D$p$) bound state will reduce to the solution of D$(p-2)$-branes
with two smeared coordinates and zero $B$ field in the decoupling limit and
the large $au >>1$ limit. The latter condition is necessary for the validity
of the dual gravity description. The worldvolume theory of the D$(p-2)$-branes
should be an ordinary gauge field theory with gauge group $U(N_{p-2})$ in
($p+1)$ dimensions if the number of the coincident D$(p-2)$-branes is
$N_{p-2}$. From the viewpoint of the thermodynamics of dual gravity solutions
for the bound states (D$(p-2)$, D$p$), we have found that $N_p$ coincident
D$p$-branes with NS $B$ field is equivalent to $N_{p-2}$ coincident
D$(p-2)$-branes with two smeared coordinates and no $B$ field. In the
equivalence, $N_p$ and $N_{p-2}$ must obey the relation (\ref{number}),
where $\tilde{V}_2$ is the area of the two additional dimensions and
$\tilde{b}$ is a noncommutativity parameter. When the volume of the torus is
sent to infinity keeping this relation, the ordinary super Yang-Mills theory
reduces to the one with gauge group $U(\infty)$ in $(p-1)$ dimensions. We
have identified the Yang-Mills coupling constant for the ($p-1)$-dimensional
ordinary Yang-Mills theory.

We have also shown the equivalence from the thermodynamics and dynamics
of two probes in the background of the bound states (D$(p-2)$, D$p$). One
of the probes is a bound state of D$p$- and D$(p-2)$-branes, which we called
a noncommutative D$p$-brane probe. The other is a D$(p-2)$-brane probe.
In the asymptotically flat limit, when the two probes fall into the horizon
of the source from spatial infinity, their static interaction potentials
at the horizon  are converted into heat and thereby are absorbed by the
source. In this process, the first law of black hole thermodynamics is
obeyed. In the decoupling limit, we
have found that the thermodynamics and dynamics of the two probes are
identical if the numbers of probe branes satisfy the relation (\ref{3e18}),
completely the same relation as (\ref{number}). As a byproduct, we have
found that the free energy of the noncommutative D$p$-brane probe in the
D$p$-brane background with a nonvanishing NS $B$ field is the same as that
of a D$p$-brane probe in the D$p$-brane background without $B$ field. It shows
that the thermodynamics of the noncommutative super Yang-Mills coincides with
the ordinary case in the large $N$ limit, not only in the Higgs branch, but
also in the Coulomb branch. In addition, from the analysis of dynamics of
probes, we have derived that the non-extremal D$p$-branes can be located at
the horizon. Our discussions support the argument by Lu and Roy \cite{Lu}
that there is  an equivalence between the noncommutative super
Yang-Mills with gauge group $U(N_p)$ in ($p+1)$ dimensions with two dimensions
compactified on a torus and the ordinary one with gauge group
$U(N_{p-2})$ in ($p-1$) dimensions when the area of the torus $\tilde{V}_2
\rightarrow \infty$, with the relation (\ref{number}) between $N_p$ and
$N_{p-2}$. This result is also consistent with the Morita
equivalence~\cite{Hashimoto2}.

\section*{Acknowledgments}

We would like to thank J.X. Lu and E. Kiritsis for helpful correspondences.
This work was supported in part by the Japan Society for the Promotion of
Science and by grant-in-aid from the Ministry of Education, Science,
Sports and Culture No. 99020.

\newpage

\end{document}